\documentclass[prb,twocolumn,aps,superscriptaddress,showpacs]{revtex4-2}

\usepackage{amsmath,amssymb}
\usepackage{graphicx}
\usepackage{xcolor}
\usepackage{graphicx}
\DeclareGraphicsExtensions{.pdf,.eps,.png,.jpg} \graphicspath{{./figs/}}
\usepackage{dcolumn}
\usepackage{bm}
\usepackage{tabularx}
\usepackage{upgreek}
\usepackage{multirow}
\usepackage{epstopdf}
\usepackage{gensymb}

\usepackage{amsmath}%
\usepackage{amsfonts}%
\usepackage{amssymb}%
\usepackage[latin1]{inputenc}
\usepackage{graphicx}
\usepackage{epstopdf}
\usepackage{color}
\usepackage{ulem}
\usepackage[english]{babel}
\usepackage{natbib}
\usepackage[colorlinks=true, citecolor=blue, linkcolor=blue, urlcolor=blue]{hyperref}
\usepackage{placeins}
\usepackage{booktabs}
\usepackage{enumitem}
\usepackage{subfigure}
\usepackage{multirow}
\usepackage{inputenc}
\usepackage{braket}
\usepackage{svg}
\usepackage{dcolumn}
\usepackage{bm}
\usepackage{siunitx}
\usepackage{mathtools}
\usepackage{soul}

\begin{document}

\title{Quasi-two-dimensional ferromagnetism in the triangular magnet EuAl$_{12}$O$_{19}$}

\author{G. Bastien}
\email{gael.bastien@matfyz.cuni.cz}
\affiliation{Charles University, Faculty of Mathematics and Physics, Department of Condensed Matter Physics, Prague, Czech Republic}
\author{Q. Courtade}
\affiliation{Charles University, Faculty of Mathematics and Physics, Department of Condensed Matter Physics, Prague, Czech Republic}
\author{A. Eli\'a\v{s}}
\affiliation{Charles University, Faculty of Mathematics and Physics, Department of Condensed Matter Physics, Prague, Czech Republic}
\author{T. Haidamak}
\affiliation{Charles University, Faculty of Mathematics and Physics, Department of Condensed Matter Physics, Prague, Czech Republic}
\author{P. Proschek}
\affiliation{Charles University, Faculty of Mathematics and Physics, Department of Condensed Matter Physics, Prague, Czech Republic}
\author{M. Du\v{s}ek}
\affiliation{FZU-Institute of Physics, Czech Academy of Sciences, Na Slovance 2, 182 00 Prague, Czech Republic}
\author{J. Priessnitz}
\affiliation{Charles University, Faculty of Mathematics and Physics, Department of Condensed Matter Physics, Prague, Czech Republic}
\affiliation{IT4 Innovations, V\v{S}B - Technical University of Ostrava, 17. listopadu 2172/15, 708 00 Ostrava-Poruba, Czech Republic}
\author{P. Bal\'a\v{z}}
\affiliation{FZU-Institute of Physics, Czech Academy of Sciences, Na Slovance 2, 182 00 Prague, Czech Republic}
\author{R. H. Colman}
\affiliation{Charles University, Faculty of Mathematics and Physics, Department of Condensed Matter Physics, Prague, Czech Republic}

\date{\today}

\begin{abstract}
We report ferromagnetic ordering at $T_\mathrm {C}=1.3\,$K in the quasi-two-dimensional magnet EuAl$_{12}$O$_{19}$ with large spins $S=7/2$. This ferromagnetic state was characterized by magnetization and specific heat measurements and the experimental results were compared with classical Monte Carlo simulations. They reveal a strong single ion anisotropy leading to an uniaxial spontaneous magnetization along the $c$ axis. Furthermore the application of a magnetic field in the hard magnetization plane $ab$ reduces the Curie temperature down to $T=0.5\,$K at $\mu_\mathrm{0} H \approx 0.4\,$T while preserving the second order nature of the ferromagnetic transition.

\end{abstract}


\maketitle

\section{Introduction}

Ferromagnetism at the quasi-two-dimensional limit has attracted much attention, since its study in bulk material will support the development of magnetic monolayers with industrial application. 
Up to now, only a few layered semiconducting or insulating ferromagnets have been reported such as the triangular lattice ferromagnets AB$_2$M(VO$_4$)$_2$ (A = Ba, Sr,  B = Ag, Na, M = Co, Ni)~\cite{Moeller2012, Nakayama2013}, the honeycomb ferromagnets BaFe$_2$(PO$_4$)$_2$~\cite{Kabbour2012}, CrI$_3$~\cite{McGuire2015}, VI$_3$~\cite{Son2019, Dolezal2019}, Cr$_2$X$_2$Te$_2$ (X = Si, Ge)~\cite{Carteaux1995, Liu2016, Lin2017} and the Kagome ferromagnet Li$_9$Cr$_3$(P$_2$O$_7$)$_3$(PO$_4$)$_2$~\cite{Magar2022}. The most studied are  the honeycomb ferromagnets CrI$_3$, VI$_3$, Cr$_2$X$_2$Te$_2$ (X = Si, Ge), since they are Van der Waals materials, which can be exfoliated down to the single-layer limit~\cite{Huang2017}. In the case of CrI$_3$, it was even shown that the ferromagnetism survives to the monolayer limit~\cite{Huang2017}.

Single crystal magnetic studies have revealed that in most of these layered ferromagnets, a strong single ion anisotropy favors the alignment of magnetic moments along the $c$ axis i.e. transverse to the layers~\cite{Nakayama2013, McGuire2015, Williams2015, Liu2019}. This anisotropy is a key ingredient for the stabilization of ferromagnetism down to the single-layer limit~\cite{Mermin1966, Carteaux1995, Williams2015}.
While ferromagnetic transitions usually evolve rapidly to a crossover under the application of magnetic field, these anisotropic ferromagnets show a slightly different behavior under the application of magnetic field in the hard magnetization plane $ab$~\cite{Selter2020}. Indeed, the
ferromagnetic transition shifts to lower temperature until a triple point where it evolves into two crossover lines~\cite{Selter2020, Spachmann2023, Zhang2023}.

The hexaaluminates with the formula  LnMAl$_{11}$O$_{19}$ (Ln = La-Tb, M = Mg, Zn) are triangular magnets close to the two-dimensional limit~\cite{Ashtar2019a, Ashtar2019}. They harbor the magnetoplumbite crystal structure where the magnetic layers are well-separated by non-magnetic spinel blocks~\cite{Ashtar2019a, Holtstam2020} (See Fig.~\ref{cell}). They are triangular lattice antiferromagnets and they were proposed as interesting materials for the study of quantum magnetism due to their magnetic frustration~\cite{Ashtar2019a, Ni2022}. Especially, the compound PrZnAl$_{11}$O$_{19}$ has recently been proposed as a gapless quantum spin liquid~\cite{Bu2022}.

We consider here the case of the hexaaluminate EuAl$_{12}$O$_{19}$. This compound was first reported in 1974~\cite{Verstegen1974} and its characterization was focused on its optical properties~\cite{Verstegen1974, Lee2016}. Recently, unusual dielectric properties in this material pointed out the realization of a frustrated antipolar phase~\cite{Bastien2024}. The magnetic ion Eu$^{2+}$  has the 4$f^7$ configuration, like Gd$^{3+}$, implying a large spin state $S = 7/2$ and vanishing orbital moment $L=0$.  However, the mixture of this spin-only state with non-zero orbital moments multiplets leads to non-vanishing spin-orbit coupling and it evidently confers single ion anisotropy to Eu$^{2+}$~\cite{Glazkov2005}. Moreover, the mixture of the 4$f$ orbitals with ligand orbitals leads to ferromagnetic interactions in many Eu$^{2+}$ based magnets~\cite{McGuire1964, Kunes2005}.

In this paper, we show that EuAl$_{12}$O$_{19}$ is a triangular lattice ferromagnet close to the two-dimensional limit with a low Curie temperature of $T_\mathrm{C}=1.3~$K. Its ferromagnetic ground state was characterized from magnetization and heat capacity measurements. They revealed a strong single ion anisotropy, which was quantified by modeling the magnetization with classical Monte Carlo simulations. We also draw the temperature-magnetic field phase diagram for magnetic field applied in the basal plane $ab$ to describe the effect of transverse magnetic fields in the anisotropic ferromagnet EuAl$_{12}$O$_{19}$.

\section{Experimental Methods}

Single crystals of EuAl$_{12}$O$_{19}$ were prepared by optical floating zone crystal growth, within a reducing atmosphere of 5\% H$_2$ in Ar. Prior to growth, the stochiometric dried binary oxides were intimately mixed by grinding, hydrostatically pressed into a cylindrical rod, and pre-reacted by sintering at 1473\,K in air. The floating zone growth was achieved at translational rate of 3\,mm/hr resulting in a dark multi-grain ingot. 

Large single crystal grains of several mm length could easily be isolated by cleavage, perpendicular to the crystallographic axis $c$. Orientation was confirmed by backscattering Laue diffraction prior to cutting and polishing of rectangular samples with dimensions around 1\,mm. These tiny single crystals are brown and transparent indicating that EuAl$_{12}$O$_{19}$ is a good insulator (Fig.\ref{cell} c). Their crystal structure at room temperature was determined using a Gemini diffractometer using a conventional sealed X-ray tube with a Mo anode and a graphite monochromator. The data were processed using CrysAlisPro 1.171.43.98a software (Rigaku Oxford Diffraction, 2023), and the structure was solved using Superflip~\cite{Palatinus2007} and refined using Jana2020~\cite{Petricek2023} without any restrictions. The data are characterized by a relatively strong extinction, which was corrected using a Gaussian model. 

Magnetization measurements above $T=1.8$\,K were performed using the MPMS magnetometer from Quantum design. These measurements were extended down to $T=0.5$\,K in the PPMS9 from Quantum design with the standard $^3$He insert using the Hall probe method~\cite{Flanders1985}. A Hall probe was installed in the vicinity of the sample such that the external magnetic field lays very close to the plane of the Hall probe and the magnetic field coming from the magnetization of the sample is approximately transverse to the Hall probe~\cite{Flanders1985}. The background signal and the nonlinearity of the Hall sensor were taken into account in the data analysis. Then the Hall signal was scaled to the absolute value of the magnetization measured at $T=1.8$\,K by the MPMS magnetometer. 

Specific heat was measured with the relaxation method using the PPMS9 from Quantum design with the standard $^3$He insert.

\section{Experimental results}

\begin{figure}
\begin{center}
\includegraphics[width=1\linewidth]{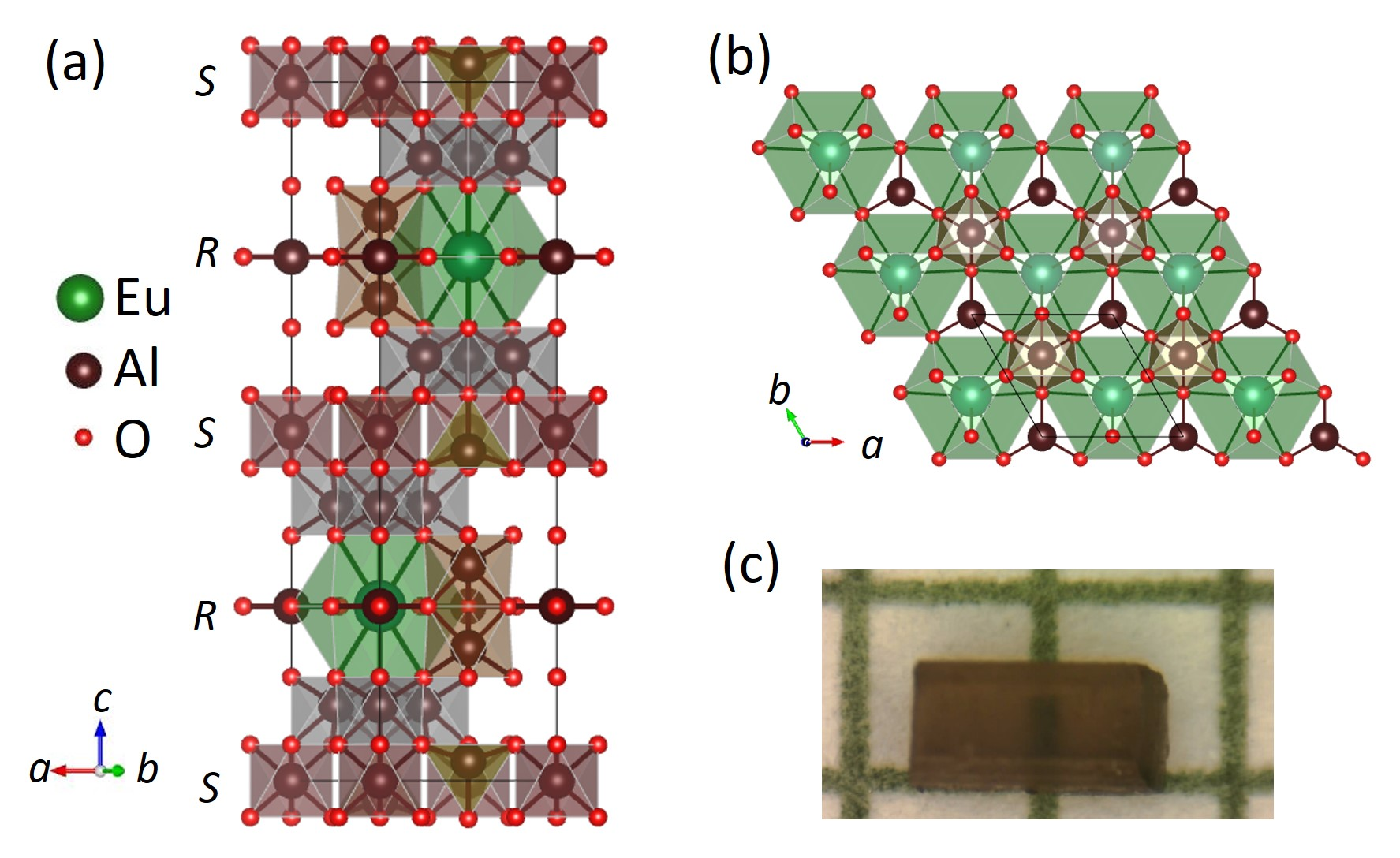}
\caption{(a) Crystal structure of EuAl$_{12}$O$_{19}$ view from the reciprocal direction $b^\star$. (b) view from the $c$ axis of the R block. Nine unit cells are represented in order to show the triangular lattice of magnetic ion Eu$^{2+}$. (c) Rectangular-shaped single crystal used for magnetization measurements on a millimeter paper.}
    \label{cell} 
\end{center}
\end{figure}

Single crystal diffraction confirms that  EuAl$_{12}$O$_{19}$ crystallizes in the magnetoplumbite structure depicted in Fig.~\ref{cell}~\cite{Verstegen1974, Holtstam2020}. This hexagonal crystal structure belongs to the space group $P6_3/mmc$ and it is named after the natural mineral PbFe$_{12}$O$_{19}$~\cite{Holtstam2020}. The complete crystallographic details are available in a cif file published as supplemental materials~\cite{sup}. The lattice parameters of EuAl$_{12}$O$_{19}$ are $a = 5.558$\,\AA\, and $c = 21.985$\,\AA . The magnetic ions Eu$^{2+}$ form planar triangular lattices, which are well separated from each other by a spinel block $S$. The magnetic sites are connected by a superexchange path via a single oxygen ion.  Interestingly this oxygen ion sits at a low symmetry position with an Eu-O-Eu angle of about 174.2$^\circ$ allowing finite Dzyaloshinskii Moriya (DM) interactions with a DM vector along the $c$ axis. DM interactions often arise in triangular magnets upon the formation of non-collinear magnetic textures accompanied by small lattice distortions and they play a crucial role for the stabilization of these non-collinear magnetic textures~\cite{Kenzelmann2007, Xiang2011}. In the case of EuAl$_{12}$O$_{19}$, DM interactions are allowed by symmetry even in the paramagnetic and ferromagnetic phases. 

The x-ray diffraction also revealed significant displacement along the $c$ axis for the Al$^{3+}$ ion located at the center of half of the triangle of Eu$^{2+}$ magnetic ions. Our recent study of the temperature dependence of the crystal structure showed that these Al$^{3+}$ ions remain disordered at any temperature due to electric dipole frustration~\cite{Bastien2024}. They induce dynamic structural disorder which might affect the magnetic properties of EuAl$_{12}$O$_{19}$.

\begin{figure}
\begin{center}
\includegraphics[width=0.9\linewidth]{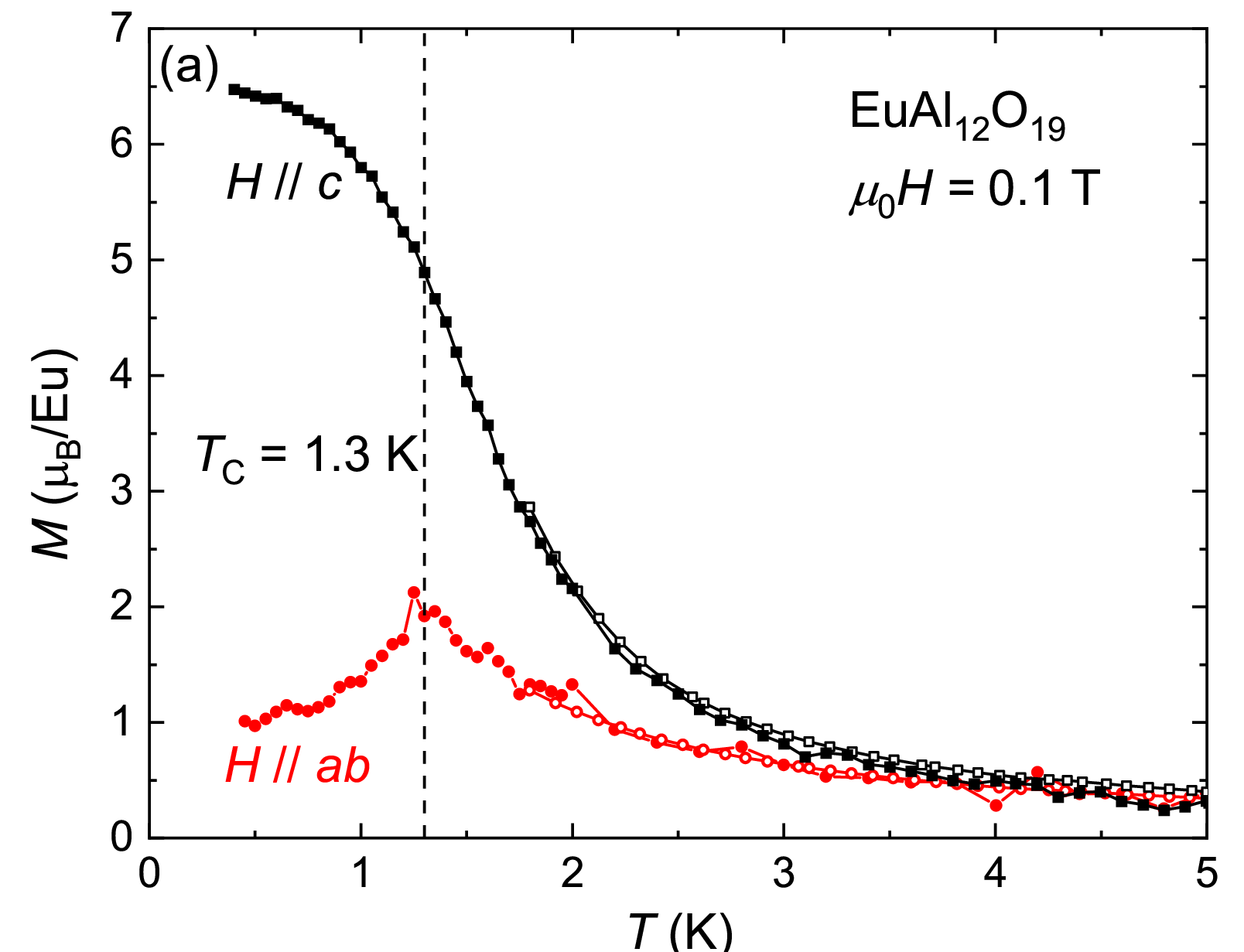}
\includegraphics[width=0.9\linewidth]{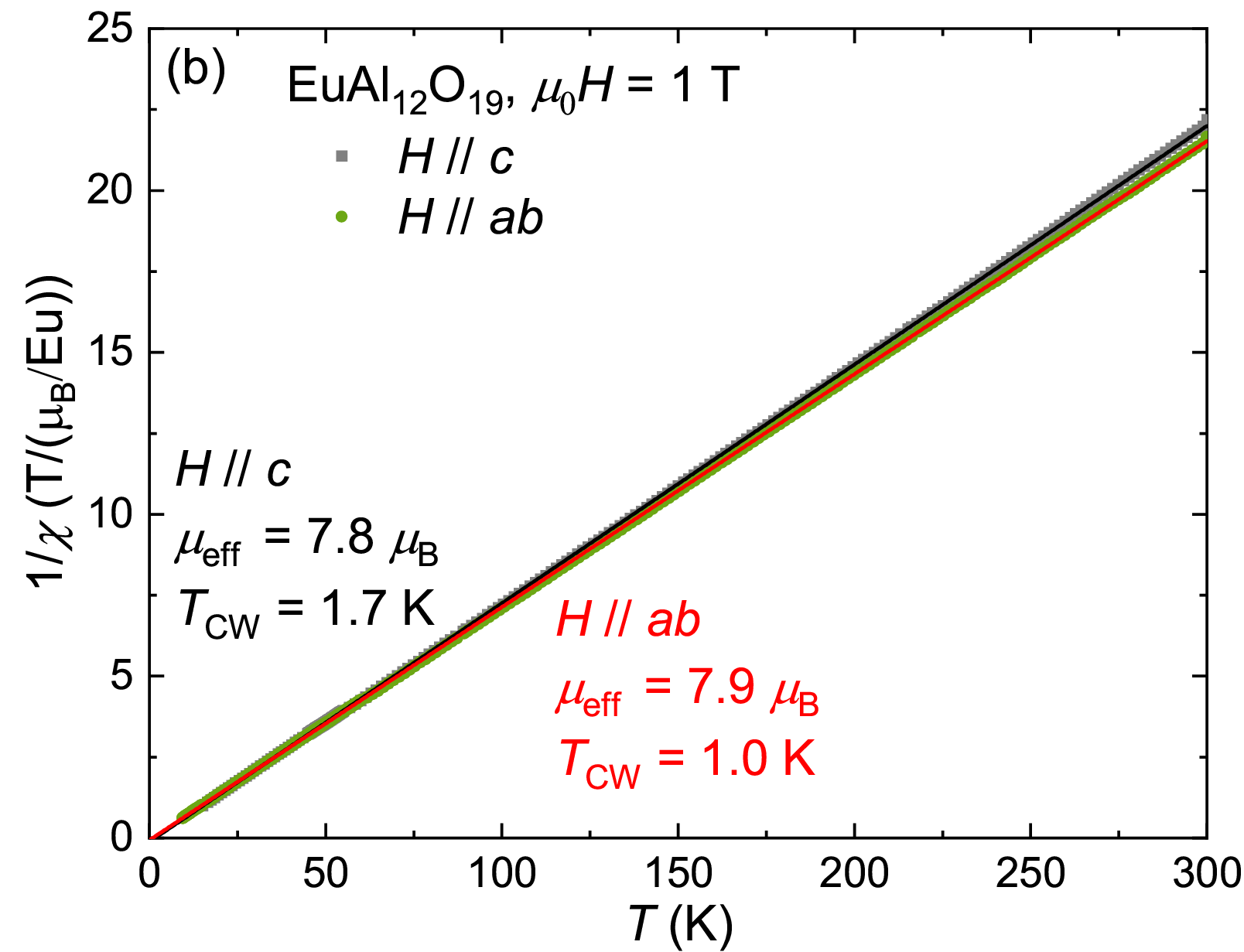}
\caption{(a) Magnetization of EuAl$_{12}$O$_{19}$ as a function of temperature under a magnetic field of $\mu_\mathrm{0} H = 0.1$\,T applied both in the $ab$ plane and along the $c$ axis. Full and open symbols correspond to data points obtained using the Hall sensor and the MPMS magnetometer, respectively. (b) Inverse magnetic susceptibility $1/\chi$ of EuAl$_{12}$O$_{19}$ as a function of temperature extracted from a magnetization measurement under a magnetic field of $\mu_\mathrm{0} H = 1$\,T. The solid lines are Curie-Weiss fits on the temperature interval 15\,K$< T <$\,300\,K.}
    \label{MvsT} 
\end{center}
\end{figure}

The magnetization of EuAl$_{12}$O$_{19}$ as a function of temperature under the external magnetic field of $\mu_\mathrm{0} H = 0.1$\,T is represented in Fig.~\ref{MvsT}(a). 
The magnetization along the easy magnetization axis $c$ increases strongly upon cooling indicating ferromagnetism. On the contrary, the magnetization measured with magnetic field applied in the $ab$ plane undergoes a sharp maximum at the Curie temperature $T_\mathrm{C} = 1.3$\,K. Such behavior is a common signature of strongly anisotropic ferromagnetism~\cite{Hardy2011, Mineev2011, Nakayama2013, Selter2020}. No splitting was observed between the magnetization measured upon cooling in zero magnetic field and the magnetization measured upon cooling under magnetic field. It shows that the ferromagnetism in EuAl$_{12}$O$_{19}$ is in the soft limit as usual in quasi-two dimensional magnets~\cite{Moeller2012, Lin2017, Selter2020}

The magnetic susceptibility in the paramagnetic state can be well fitted by the Curie-Weiss law on a large temperature interval 15\,K$<T<$\,300\,K with positive Curie-Weiss temperatures for both magnetic field directions, $T_\mathrm{CW}= 1.0\,\pm\,0.3$\,K for $H//ab$ and  $T_\mathrm{CW}= 1.7\,\pm\,0.3$\,K for $H$//$c$ (Fig.\Ref{MvsT}(b)). It confirms that the magnetic interactions between nearest neighbors are ferromagnetic and it confirms the easy axis magnetic anisotropy. These Curie-Weiss fits give effective moments of $\mu_\mathrm{eff}=7.9\,\pm\,0.1\,\mu_\mathrm{B}$ for $H//ab$ and $\mu_\mathrm{eff}=7.8\,\pm\,0.1\,\mu_\mathrm{B}$ for $H//c$. These values are similar to the expected value for a free Eu$^{2+}$ ion $\mu_\mathrm{ eff}=7.94\,\mu_\mathrm{B}$ confirming that magnetic moments are large spins $S=7/2$.

\begin{figure*}[t]
\begin{minipage}{0.49\linewidth}
\begin{center}
\includegraphics[width=1\linewidth]{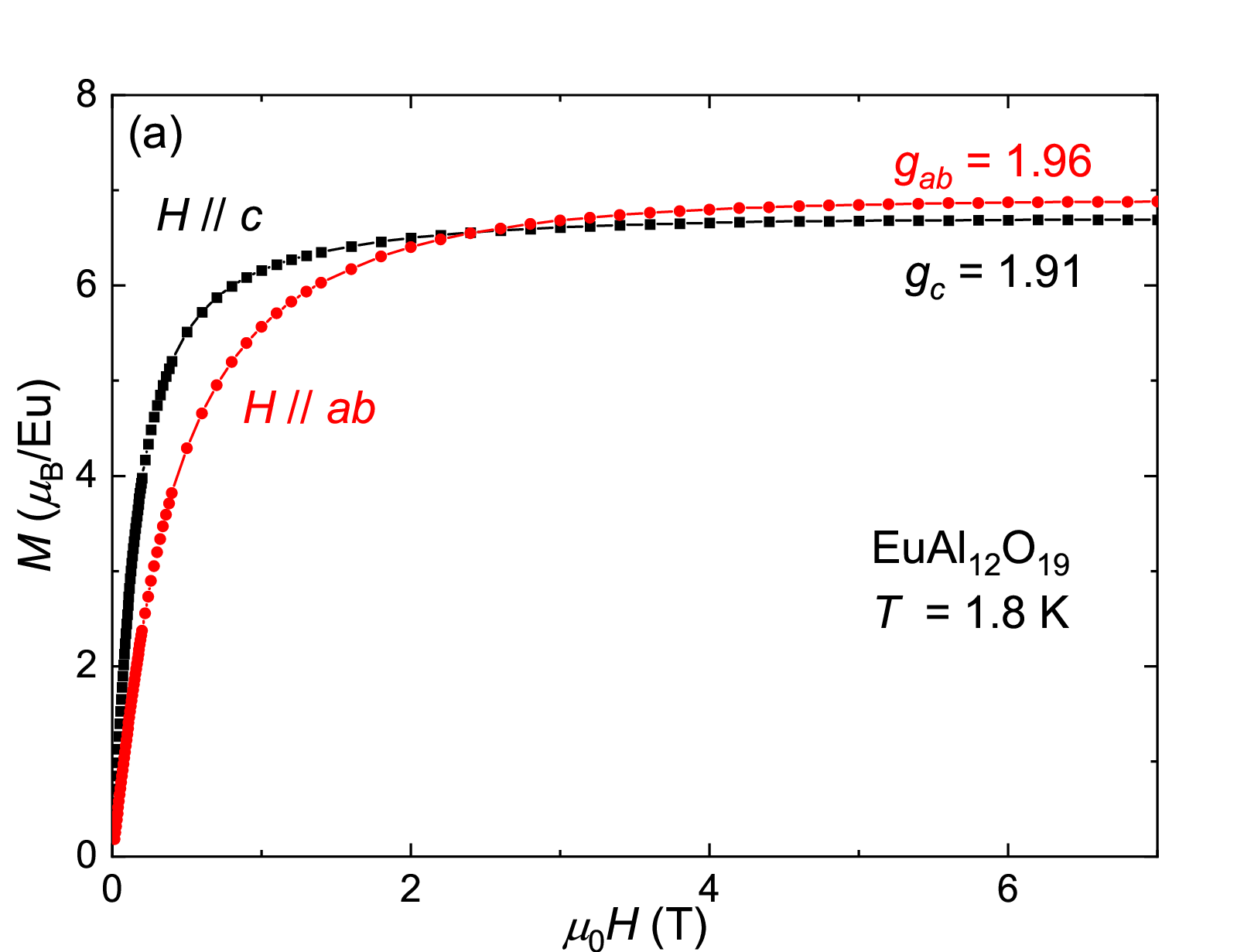}
\includegraphics[width=1\linewidth]{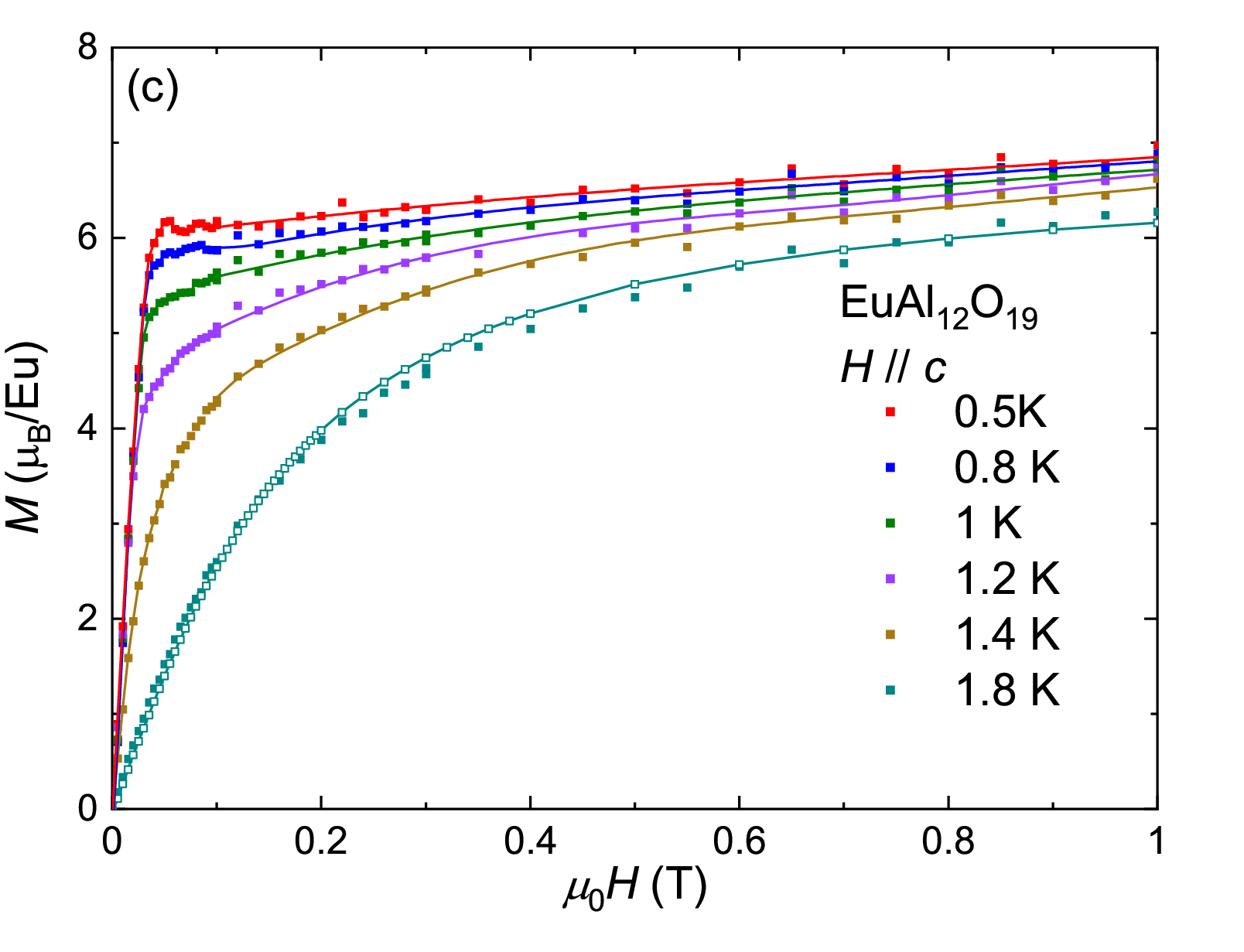}
\end{center}
\end{minipage}
\hfill
\begin{minipage}{0.49\linewidth}
\begin{center}
\includegraphics[width=1\linewidth]{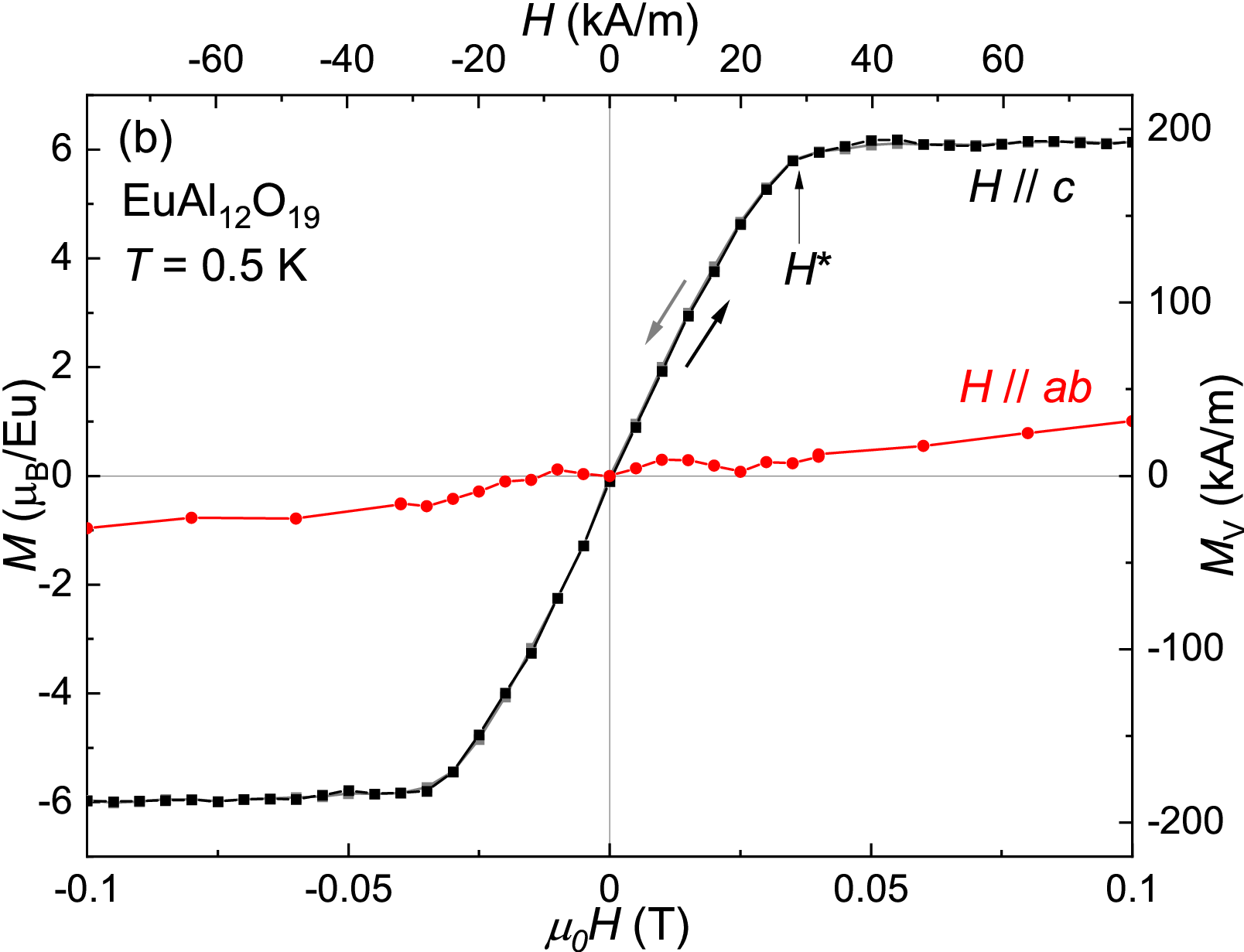}
\includegraphics[width=1\linewidth]{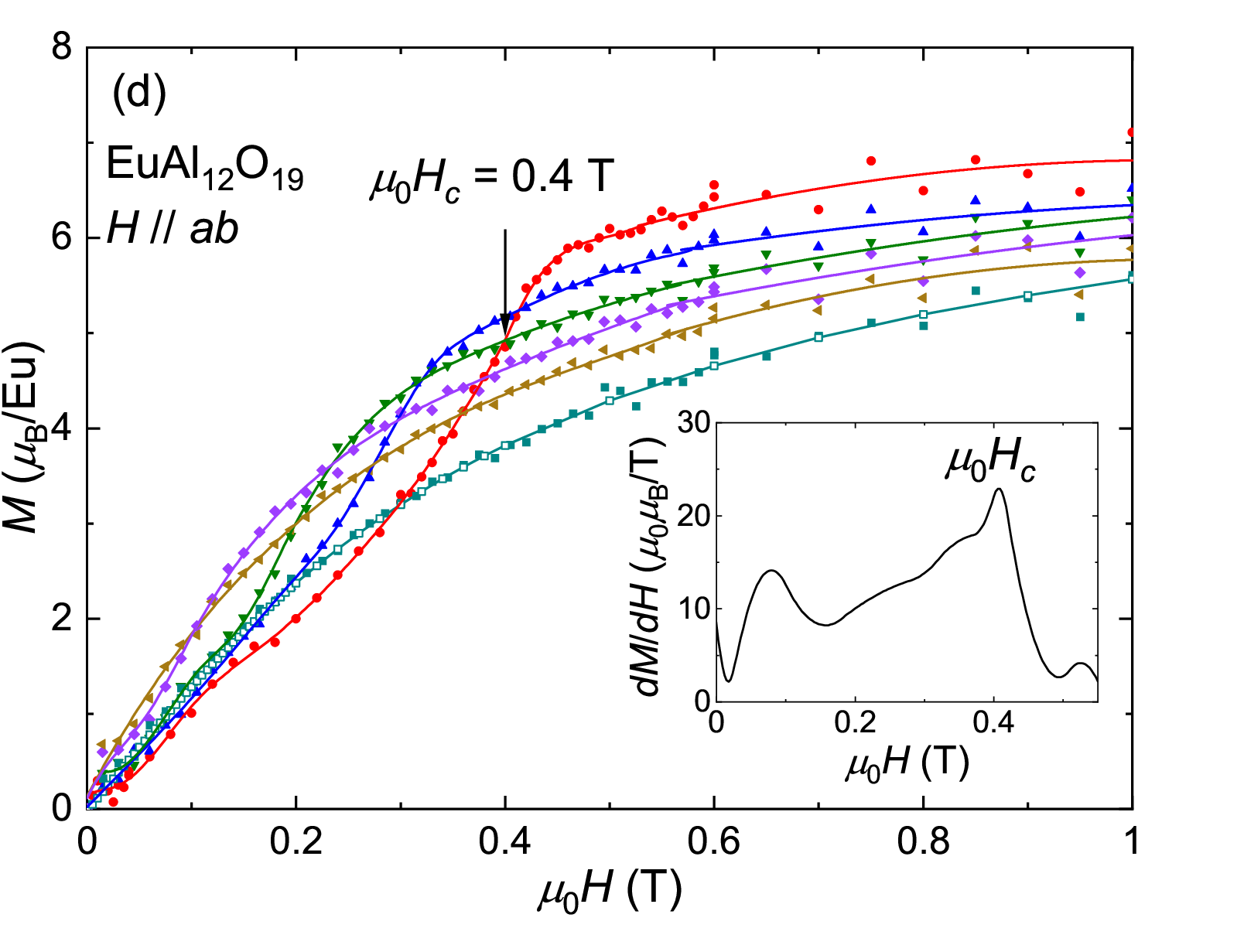}
\end{center}
\end{minipage}
\caption{
(a) Field dependence of the magnetization of EuAl$_{12}$O$_{19}$ in the paramagnetic state at $T = 1.8$\,K. The values of the magnetization at saturation give us $g$-factor values which are indicated on the graph. (b) Magnetization as a function of external magnetic field  $\mu_\mathrm{0} H$ in EuAl$_{12}$O$_{19}$ in the ferromagnetic state at $T = 0.5$\,K. No hysteresis in magnetic field is observed. The right axis indicates the magnetization per volume unit. (c-d) Magnetic field dependence of the magnetization under magnetic field applied along the $c$ axis in (c) and in the $ab$ plane in (d). Full and open symbols correspond to data points obtained using the Hall sensor and the MPMS magnetometer, respectively. The arrow in (d) indicates the magnetic field $\mu_\mathrm{0}H_c=0.4\,$T where the magnetic field-induced transition occurs for $T=0.5\,$K. The inset shows the derivative of the magnetization $dM/dH$ as a function of magnetic field.}
\label{MvsH}
\end{figure*}

The measurement of the magnetization in the paramagnetic state indicates a change of magnetic anisotropy upon cooling. Indeed, the room temperature magnetic anisotropy $\chi_c/\chi_{ab}<1$ is opposite with respect to the low temperature magnetic anisotropy $\chi_c/\chi_{ab}>1$ (Fig.~\ref{MvsT}(b)). The change of magnetic anisotropy appears also on the field dependence of the magnetization at $T= 1.8$\,K 
(Fig.~\ref{MvsH}(a)). While the magnetic susceptibility at $T=$ 1.8\,K is larger for magnetic field along the $c$ axis, the magnetization reaches saturation around $\mu_0 H = 3$\,T with a higher magnetization at saturation for magnetic field applied within the $ab$ plane than for magnetic field along the $c$ axis. This change of the magnetic anisotropy with magnetic field and temperature results from an anisotropy of the $g$ factor opposed to the anisotropy of the ferromagnetic state. The $g$ factor can be deduced from the magnetization at the saturation using the relation $ M_{sat}=gS$. Using the expected spin value of $S = 7/2$ we quantify the $g$ factor anisotropy with $g_{ab} = 1.96$ and $g_c = 1.91$.

Magnetization measurements were extended down to $T=0.5$\,K using Hall probes to characterize the magnetic ground state of EuAl$_{12}$O$_{19}$ and the results are represented in Fig~\ref{MvsH}(b-d).
The magnetization along the $c$ axis at $T=0.5\,$K as a function of external magnetic field harbors a finite slope at low magnetic field  without any hysteresis within the resolution of the correction estimate. This behavior defined as soft ferromagnetism is commonly observed in quasi-two-dimensional ferromagnets~\cite{Nakayama2013, Liu2019, Selter2020, Magar2022}. The magnetization along the $c$ axis harbors a change of slope at the external field of $H^\star=29$\,kA/m ($\mu_\mathrm{0}H^\star= 36$\,mT). To confirm that this kink corresponds to the achievement of a single magnetic domain, we compared the value of $H^\star$ with the demagnetization field $H_D=NM_V$, where $N$ and $M_V$ stand for the demagnetization factor and the volumic magnetization, respectively. Since we used rectangular shaped crystals for practical reasons, the demagnetization factor is inhomogeneous~\cite{Bahl2021}. However, the average value of the demagnetization factor can be estimated from the equation in Ref.~\cite{Aharoni1998} at $N_c=0.17$ implying a demagnetization field of $H_D=31$\,kA/m at $H^\star=29$\,kA/m. The proximity of the values of $H^\star$ and $H_D$ confirm that below $H^\star$, the demagnetization field compensates the external field. As a consequence, the finite slope in the field dependence of the magnetization below $H^\star$ can be ascribed to the continuous evolution of the domain populations under magnetic field. On the contrary, under magnetic field applied in the $ab$ plane, the average demagnetization field $H_D=NM_V$ ($N_{ab}=0.09$) remains much weaker than the external field $H$ implying an absence of ferromagnetic component in this direction. These results show that the ferromagnetism in EuAl$_{12}$O$_{19}$ is close to the soft limit and uniaxial with magnetic moments pointing along the $c$ axis.

Magnetization measurements in the basal plane $ab$ revealed a field-induced magnetic transition at $\mu_\mathrm{0}H_c=0.4$\,T for $T=0.5\,$K indicated by a maximum of the slope $dM/dH$. This transition corresponds to the reorientation of the magnetization from the $c$ axis toward the direction of the applied magnetic field. The magnetic field $\mu_\mathrm{0}H_c=0.4$\,T where this transition occurs is relatively high compared to a Curie temperature of $T_\mathrm{C}=  1.3$\,K, implying a strong magnetic anisotropy. This strong magnetic anisotropy is an interesting property of EuAl$_{12}$O$_{19}$ since it is one of the key ingredients for the stabilization of ferromagnetic order at the quasi-two-dimensional limit~\cite{Mermin1966}.

\begin{figure}
\begin{center}
\includegraphics[width=0.9\linewidth]{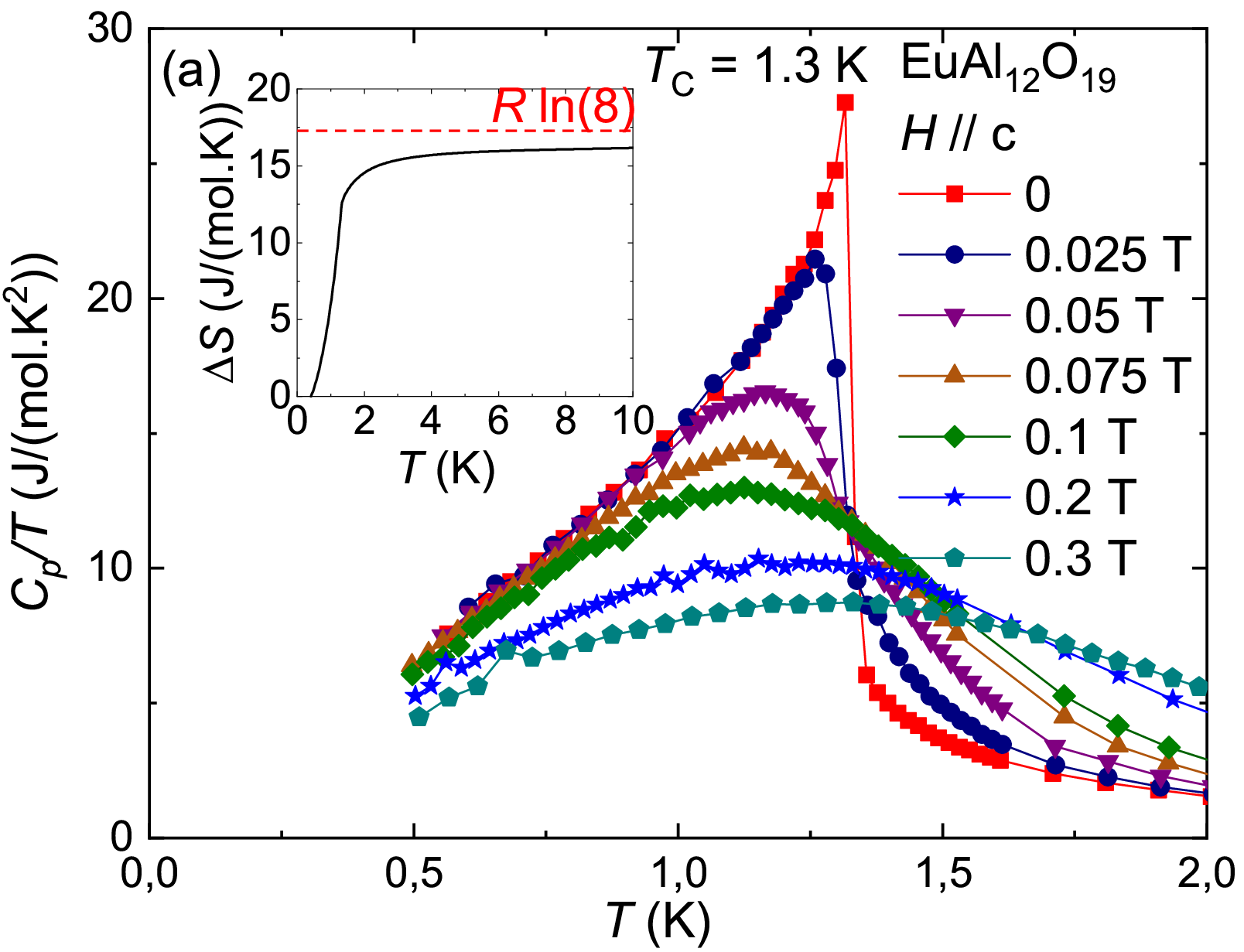}
\includegraphics[width=0.9\linewidth]{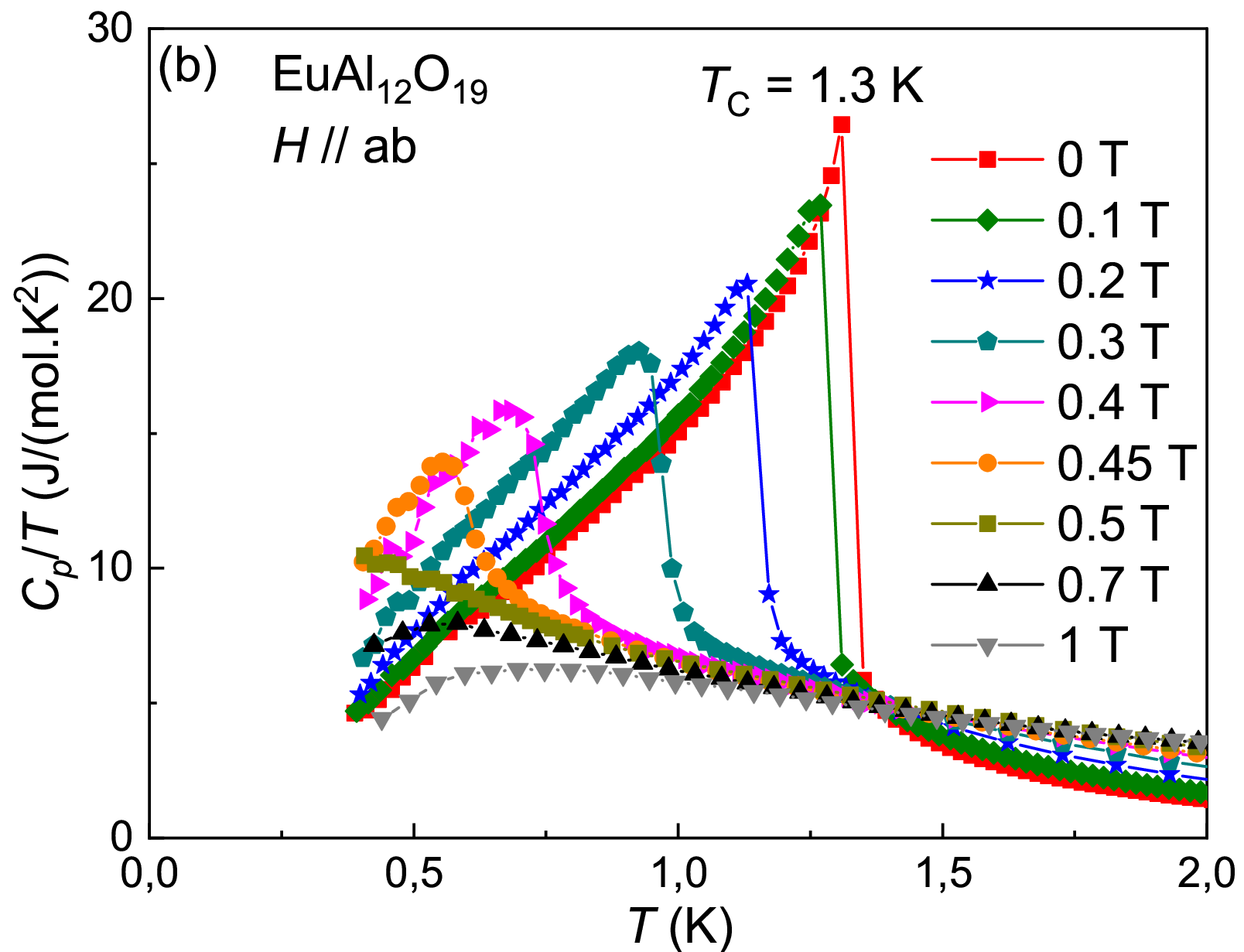}
\caption{(a) Specific heat of EuAl$_{12}$O$_{19}$  as a function of magnetic field applied along the $c$ axis. The average demagnetization factor of the rectangular crystal was estimated at $N\approx0.29$. The inset shows the entropy variation between 0.4\,K and 10\,K in absence of external magnetic field. The dashed line indicates the expected entropy change of $\Delta S=R\, {\mathrm  ln}$(8). (b) Same figure for magnetic field applied within the basal plane $ab$. 
The average demagnetization geometrical factor for this field direction was estimated at $N\approx0.30$.}
\label{Cp}
\end{center}
\end{figure}

For more insights into the magnetic ground state of EuAl$_{12}$O$_{19}$ and the magnetic field-induced transition for $H//ab$, we performed specific heat measurements and $C_p/T$ as a function of $T$ is plotted in Fig.~\ref{Cp}. In the absence of magnetic fields it shows a clear second order magnetic transition at the Curie temperature at $T_\mathrm{C} = 1.3$\,K. The integration of $C_p/T$ gives a change of magnetic entropy between $T=0.4\,$K and $T=10\,$K of about 16\,J/(mol.K). 
Considering that the peak was not fully integrated due to absence of data below 0.4\,K,
This value agrees well with the expectation for a $S=7/2$ system of an entropy  change of $\Delta S=R\,\mathrm{ln}(8)=17.3$\,J/(mol.K).

Under magnetic field applied along the easy magnetization axis $c$ the second order magnetic transition evolves towards a crossover as expected for a ferromagnetic transition. On the contrary, the magnetic transition remains second order under magnetic field applied within the $ab$ plane at least up to an external magnetic field of $\mu_\mathrm{0} H=0.45\,$T. It confirms that the field-induced magnetic transition observed in the magnetization at $H_c$ can be interpreted as a field-induced magnetic transition from a ferromagnetic state towards the paramagnetic state. This behavior is unusual in ferromagnets~\cite{Selter2020} and it likely comes from the competition between the strong magnetic anisotropy and ferromagnetic interactions. Magnetic field-induced second order transitions are usually observed in antiferromagnets so it might also indicate a slight antiferromagnetic component in the $ab$ plane in addition to the large spontaneous magnetization along the $c$ axis i.e. a canted ferromagnetic state.  At $\mu_\mathrm{0} H=0.5\,$T $C_p/T$ increases upon cooling down to the lowest temperature measured, $T= 0.4$\,K, leaving the possibility of a field-induced magnetic quantum critical point at this magnetic field. Under a magnetic field of $\mu_\mathrm{0} H=0.7\,$T or higher, $C_p/T$ as a function of temperature $T$ undergoes a broad maximum indicating a continuous evolution from magnetic saturation at $T=0$ to a weakly polarized state at high temperature.

\begin{figure}
\begin{center}
\includegraphics[width=0.9\linewidth]{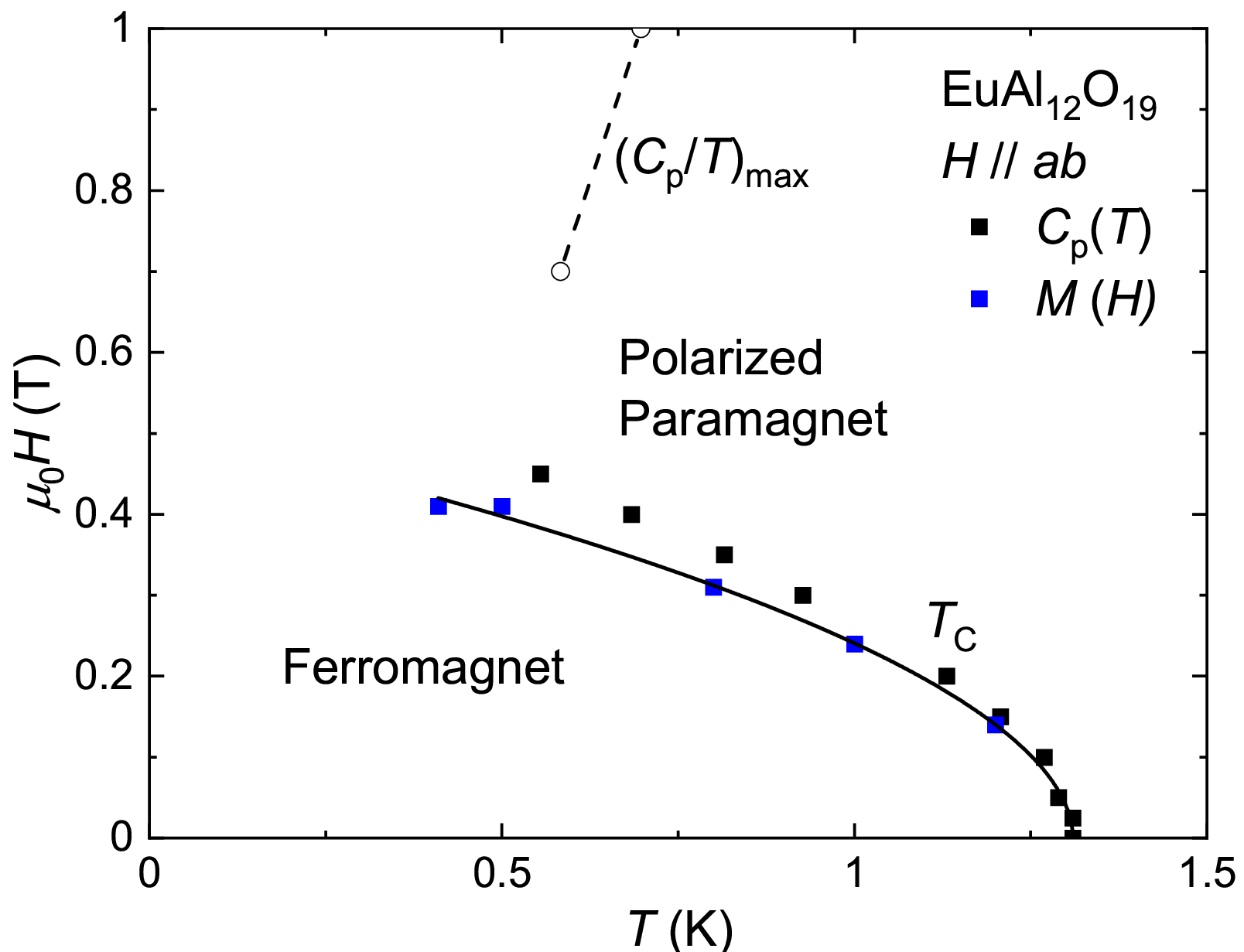}
\caption{Magnetic field - temperature ($H-T$) phase diagram of  EuAl$_{12}$O$_{19}$ under magnetic field applied within the hard magnetization plane $ab$. The solid line indicates the second order phase transition from the ferromagnetic state to the polarized paramagnet. The dashed line indicates the broad maximum observed in the temperature dependence of $C_p/T$ for $\mu_\mathrm{0}H>0.5$\,T.}
\label{BT}
\end{center}
\end{figure}

The magnetic phase diagram of EuAl$_{12}$O$_{19}$ under magnetic field applied in the $ab$ plane is represented in Fig.~\ref{BT}. The second order magnetic transition line separates the ferromagnetic state from the polarized paramagnetic state. While ferromagnets and polarized paramagnets are often considered as similar phases,
they differ in the case $H//ab$  by the direction of the magnetic moments. Indeed, in the ferromagnetic state the crystal is divided
in two domains with positive and negative projection of the magnetization on the c axis. On the contrary,
in the polarized paramagnet the magnetization is aligned with the magnetic field in the $ab$ plane.
 
 Magnetic transitions induced by a magnetic field transverse to the easy magnetization axis have previously been identified and studied in few other anisotropic ferromagnets~\cite{Levy2007, Gourgout2016, Nakayama2013, McGuire2015, Liu2019, Zhang2023, Selter2020, Spachmann2023, Zhang2023, Bitko1996, Coldea2010}.  In the intermetallic ferromagnet URhGe, the magnetic transition evolves into a first order magnetic transition at a tricritical point~\cite{Levy2007, Gourgout2016}. In the semiconducting ferromagnets Cr$_2$Si$_2$Te$_6$ and Cr$_2$Ge$_2$Te$_6$, it evolves into a crossover at a finite magnetic field~\cite{Selter2020, Spachmann2023, Zhang2023}. On the contrary, in EuAl$_{12}$O$_{19}$ the ferromagnetic transition remains of second order under magnetic fields at least until it reaches the lowest temperature of our study $T=0.4$\,K implying the possible realization of a ferromagnetic quantum critical point. Such field-induced ferromagnetic quantum critical points were previously reported in few anisotropic ferromagnets such as the dipolar ferromagnet LiHoF$_4$~\cite{Bitko1996} and the ferromagnetic spin chain CoNb$_2$O$_6$~\cite{Coldea2010}.

\begin{figure*}[t]
\begin{minipage}{0.49\linewidth}
\begin{center}
\includegraphics[width=1\linewidth]{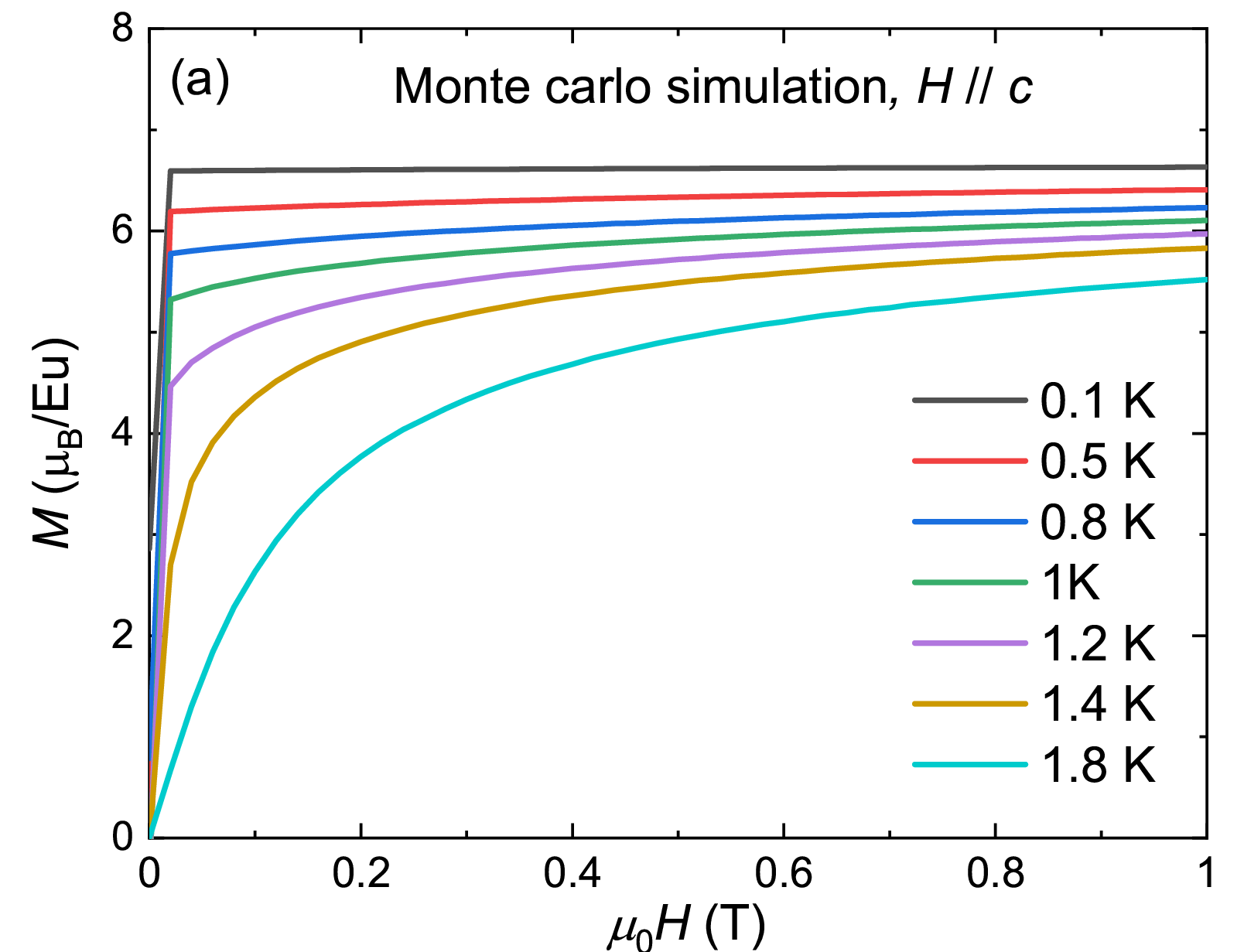}
\includegraphics[width=1\linewidth]{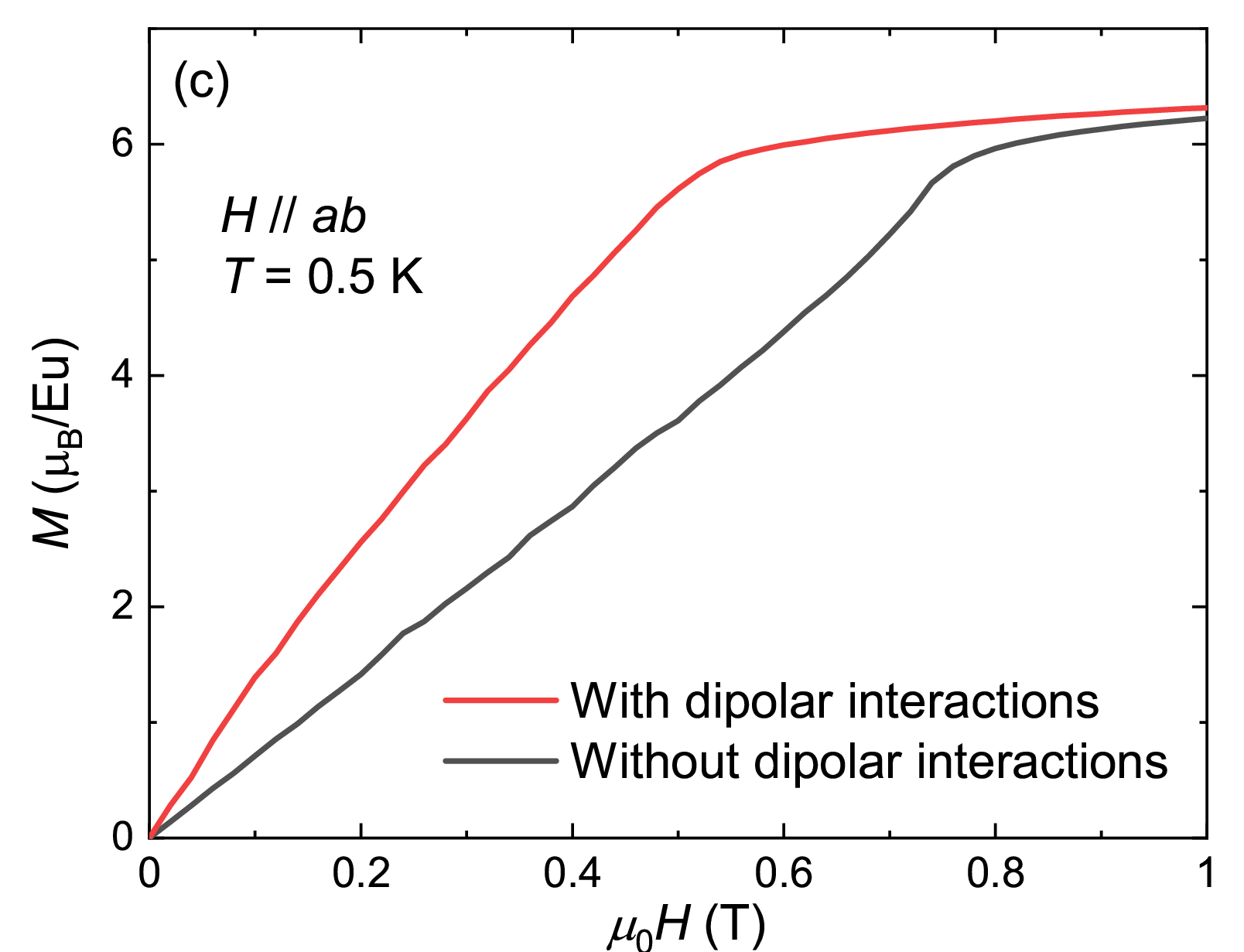}
\end{center}
\end{minipage}
\hfill
\begin{minipage}{0.49\linewidth}
\begin{center}
\includegraphics[width=1\linewidth]{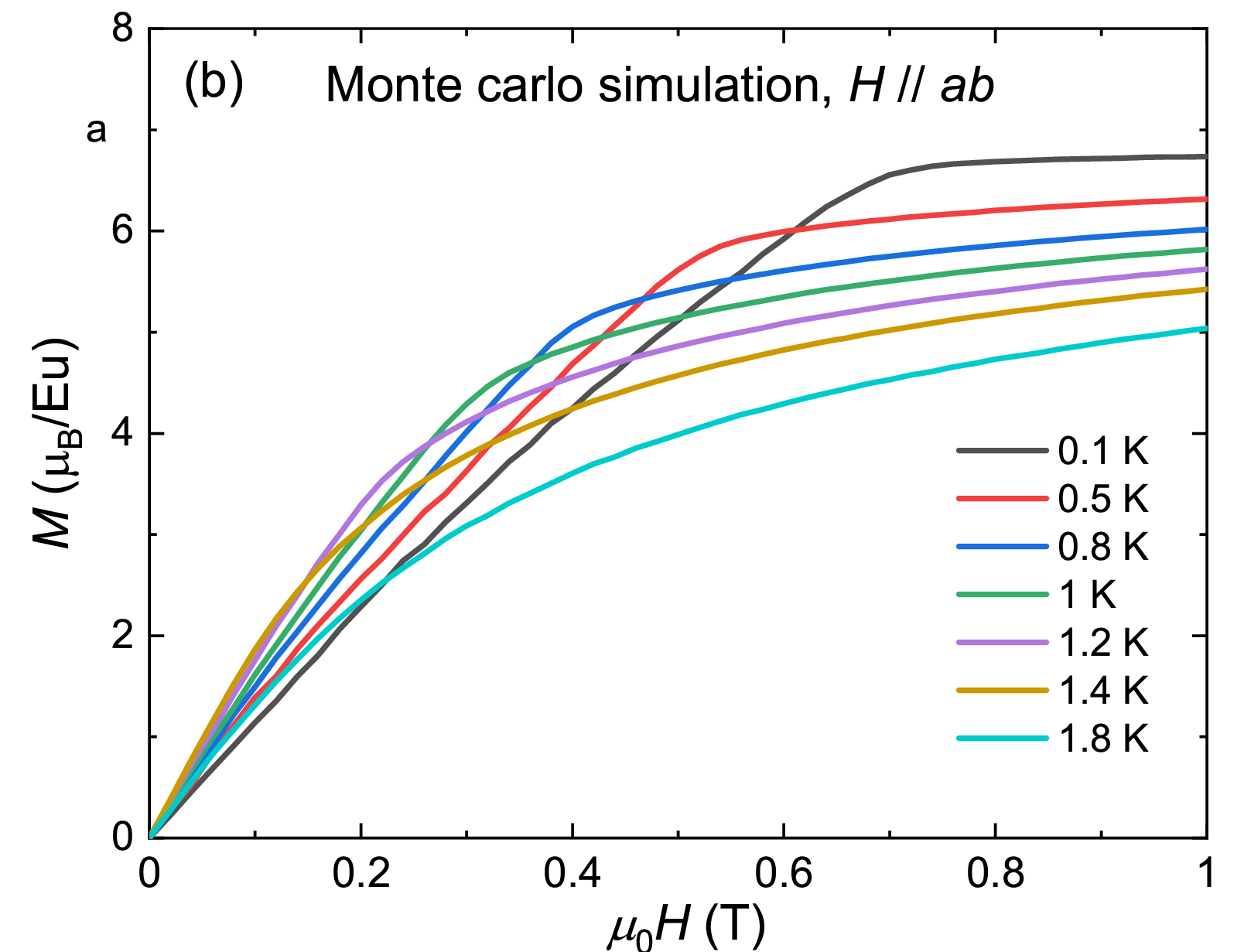}
\includegraphics[width=1\linewidth]{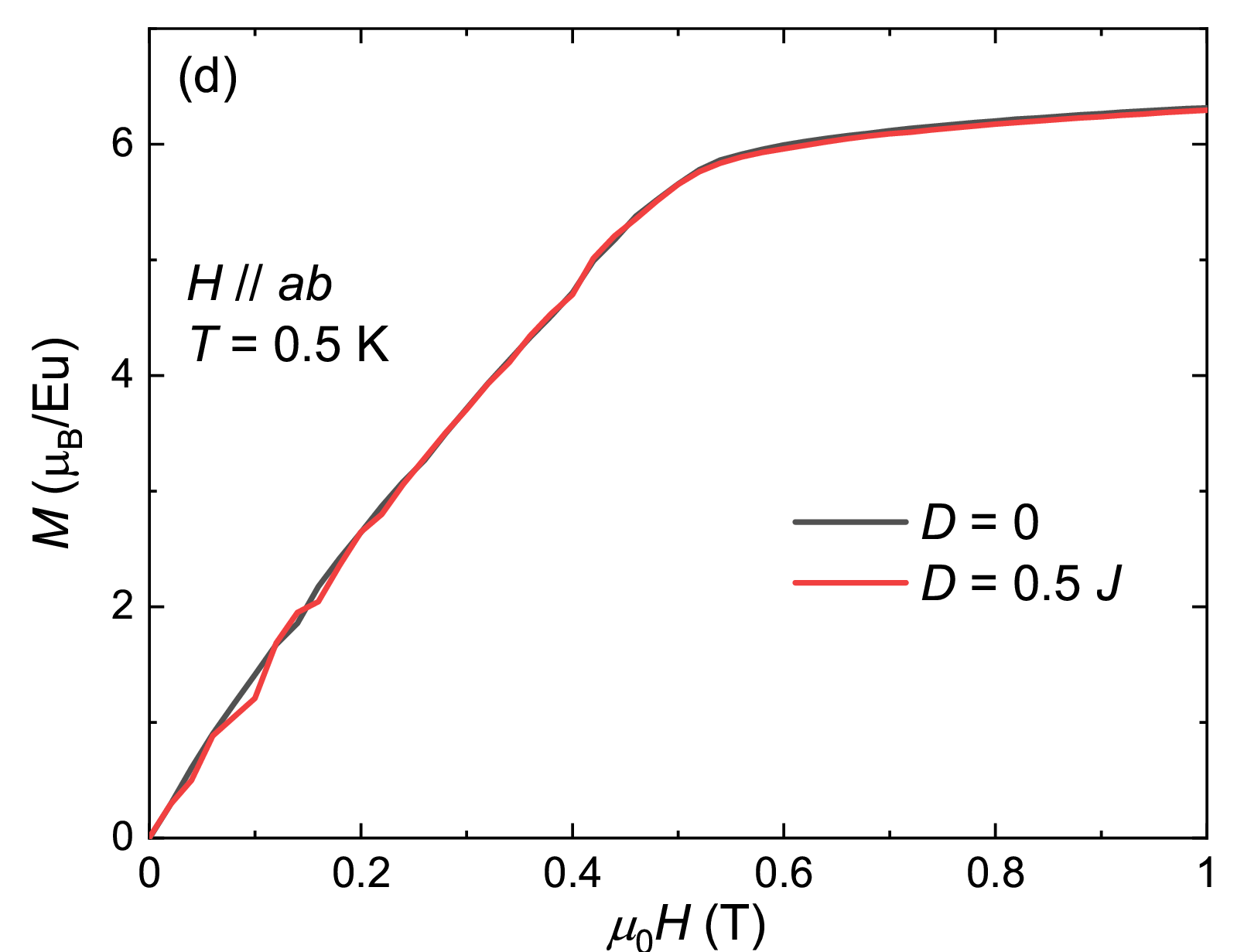}
\end{center}
\end{minipage}
\caption{(a-b) Isothermal magnetization along the $c$ axis in (a) and in the $ab$ plane in (b) computed by classical Monte Carlo calculations from the Hamiltonian~\ref{H} with parameters $J= 2.8\,\upmu eV$ and $K = 16.7\,\upmu eV$, $D=0$. These parameters were chosen for a close similarity of the results with experimental data reported in Fig.~\ref{MvsH}. (c-d) The magnetization in the $ab$ plane at $T=0.5\,$K is compared with a calculation performed upon switching off the dipolar interactions in (c) and with a calculation in presence of a strong DM interaction $D=0.5J$ in (d).}
\label{calc}
\end{figure*}

\section{Magnetic model and classical Monte Carlo Calculations}

For a quantitative analysis of the magnetic anisotropy of the ferromagnet EuAl$_{12}$O$_{19}$, we performed classical Monte Carlo calculations. The magnetism of EuAl$_{12}$O$_{19}$ was modeled by the following Hamiltonian which includes three anisotropic terms: the single ion anisotropy~\cite{Glazkov2005}, the dipolar interactions and the DM interactions:
\begin{gather}
H=\sum_{<i,j>}{JS_i.S_j}+H_{si}+H_{dip}+H_{DM}\\
H_{si}=-\sum_{i}{K(S_{i}^z)^2}\\
H_{dip}=\frac{\mu_\mathrm{0} g^2 \mu_\mathrm{B}^2}{4 \pi }\sum_{i<j}{\frac{1}{r_{ij}^3}(S_i.S_j-3(e_{ij}.S_i)(e_{ij}.S_j))}\\
H_{DM}=\sum_{<i,j>}{D_{ij}.(S_i \times S_j)}
\label{H}
\end{gather}
The directions $\hat{x}, \hat{y}$ and $\hat{z}$ are along the crystallographic and reciprocal axis $a$, $b^\star$ and $c$, respectively. In the dipolar interaction term, $e_{ij}$ and $r_{ij}$ stand for the unitary vector from the site $i$ to the site $j$ and the distance between the sites $i$ and $j$, respectively. While exchange interactions are only considered between nearest neighbors, dipolar interactions are taken into account for each pair of spins to account for long-range dipolar interactions~\cite{Politi2006}.
The presence of a single ion anisotropy in 4$f^7$-based magnets is made possible by the mixing of the spin-only ground state with multiplets of higher spectral terms~\cite{Glazkov2005}. The room temperature crystal structure of EuAl$_{12}$O$_{19}$ allows only the $c$ axis component of the DM vector $D_{ij}$.

Monte Carlo calculations were conducted with the Metropolis-Hastings algorithm using UppASD code~\cite{Skubic2008}. The $g$-factor value was taken from experiments and its slight anisotropy was neglected. The simulation domain consists of $28 \times 28 \times 7$ unit cells, each containing 2 Eu ions. We used periodic boundary conditions in all three directions. This system size is large enough to accurately grasp all featured interactions, specifically ensuring that long-rangeeffect of the dipolar interaction was accounted for. We validated this by comparing calculated magnetization vs. field dependence for multiple system sizes.
The initial spin configuration of each simulation was completely random. The measurement phase consisted of 20,000 Monte Carlo steps, preceded by an initial simulated annealing phase. The initial phase consisted of several temperature steps, gradually decreasing the temperature from $T_{init}$ to $T_{fin}$ such that $T_{i+1} = T_i /1.2$ and $T_{init} = 3.9~\mathrm{K} \approx 3\,T_\mathrm{C}$. Each temperature step consisted of 1,000 Monte Carlo steps.



The magnetic field and temperature dependence of the magnetization was computed for various values of $J$ and $K$ and the best fit of the experimental data was achieved with the values $J= 2.8\,\upmu eV$ and $K = 16.7\,\upmu eV$. The temperature and field dependence of the magnetization obtained with these parameter values and $D=0$ is represented in Fig.~\ref{calc}(a-b). Calculated magnetization curves for a large set of values of $J$ and $K$ are available on the repository Zenodo~\cite{Zenodo}.
The good agreement with experimental data confirms that a strong single ion anisotropy is needed to explain the observed magnetic anisotropy in EuAl$_{12}$O$_{19}$. The single ion anisotropy in EuAl$_{12}$O$_{19}$ strongly exceeds the strength of the magnetic interaction with $K/J \approx 6$. Single ion anisotropy comparable with the magnetic interactions were previously reported in other 4$f^7$ magnets such as the pyrochlore magnets Gd$_2$M$_2$O$_7$ (M = Ti, Sn, Ir)~\cite{Glazkov2005, Glazkov2007, Lefrancois2019} whereas magnetism of other 4$f^7$ magnets such as the Kagome magnet Gd$_3$Mg$_2$Sb$_3$O$_{14}$ and the triangular magnet KBaGd(BO$_3$)$_2$ can well be modeled upon neglecting the single ion anisotropy~\cite{Wellm2020,Xiang2023}. In this context, the single ion anisotropy in EuAl$_{12}$O$_{19}$ appears impressively strong and the microscopic origin of this strong single ion anisotropy is beyond the scope of this work.

The dipolar magnetic interactions between nearest neighbors can be estimated at $J_\mathrm{dip}=\mu_\mathrm{0} g^2\mu_\mathrm{B}^2/4 \pi a^3=1.14\,\upmu eV$, where $a$ is the unit cell parameter. This value corresponds to a ratio of $J_\mathrm{dip}/J= 0.4$, thus the dipolar term contributes significantly to the magnetic interaction in EuAl$_{12}$O$_{19}$. To test the effect of dipolar magnetic interaction on the magnetic anisotropy, we computed the magnetization in the $ab$ plane by switching off the dipolar interactions and the results are represented in Fig.~\ref{calc}(c). It results in an enhancement of the magnetic anisotropy proving that the dipolar interactions favor an easy plane anisotropy. However, their effect is strongly overcome by the single ion anisotropy in EuAl$_{12}$O$_{19}$ with a ratio of $K/J_\mathrm{dip} \approx 15$, in contrast with the triangular magnet KBaGd(BO$_3$)$_2$ where dipolar interactions induce the easy plane anisotropy~\cite{Xiang2023}.

The calculated magnetization in the $ab$ plane shows a smooth crossover from the ferromagnetic state towards paramagnetic state in contrast with the measured magnetization in EuAl$_{12}$O$_{19}$ where a maximum of the slope $dM/dB$ indicates a second order transition. This second order behavior could come from a slight tilting of the magnetic moments in EuAl$_{12}$O$_{19}$ implying an antiferromagnetic component. In order to test whether a tilting of the magnetic moments is induced by the DM interactions, we ran a second calculation with strong DM interactions $D_{ij}=\pm\,0.5J \hat{z}$ (Fig.~\ref{calc}(d)). These DM interactions do not induce any tilting of the magnetic moments and they have only a minor effect on the magnetization in EuAl$_{12}$O$_{19}$. These results rule out the induction of a small antiferromagnetic component by the DM interaction with a DM vector along the $c$ axis. This result contrasts with the case of triangular lattice antiferromagnets where weak DM interactions play an important role for the stabilization of non-collinear magnetic structures~\cite{Kenzelmann2007, Xiang2011} or skyrmion lattices~\cite{Rosales2015, Mohylna2022}. The second order nature of the magnetic field-induced transition from ferromagnetism to paramagnetism must be protected by effects which are beyond our Monte Carlo simulations such as quantum magnetic fluctuations or magnetoelastic coupling.


The dipolar interactions between adjacent planes are considered to be the strongest interplane interactions, since the superexchange path between two magnetic sites on different planes would go via at least four oxygen atoms. The dipolar interplane interaction is estimated at $J_\mathrm{dip, inter}=\mu_\mathrm{0} g^2\mu_\mathrm{B}^2/4 \pi d^3=0.13\,\upmu eV$ where $d=11.46$\,\AA\,is the shortest distance between magnetic sites on different planes. This interplane interaction is more than one order of magnitude weaker than the intraplane interaction confirming the quasi-two-dimensional behavior of the magnetism in EuAl$_{12}$O$_{19}$. Considering that the moments orient preferably along the $c$ axis and the interplane dipolar interactions favor ferromagnetic correlations for this spin component, the interplane dipolar interaction acts to stabilize  the long-rangeferromagnetic order. 



\section{Conclusion}
We report the discovery of a triangular lattice ferromagnet EuAl$_{12}$O$_{19}$ with $T_{\mathrm C}=1.3$\,K and a uniaxial spontaneous magnetization along the $c$ axis. The large magnetic anisotropy of EuAl$_{12}$O$_{19}$ comes from a strong single ion anisotropy which overcomes the dipolar interactions and stabilizes
ferromagnetism 
close to the two-dimensional limit. Under the application of a magnetic field in the $ab$ plane, the magnetic transition remains of second order and it shifts to lower temperature. It approaches zero temperature implying the possible realization of field-induced magnetic quantum criticality around $\mu_\mathrm{0} H_c \approx 0.5\,$T, which will be the scope of future studies based on measurements at temperature lower than $T=0.4$\,K.

\begin{acknowledgments}
We acknowledge funding from the Charles University in Prague within the Primus research program with grant number PRIMUS/22/SCI/016. J.P. acknowledges grant No. 22-35410K by Czech Science Foundation. Crystal growth, structural analysis and magnetic properties measurements were carried out in the MGML (http://mgml.eu/), which is supported within the program of Czech Research Infrastructures (project no. LM2023065). Computational resources were provided by the e-INFRA CZ project (ID:90254), supported by the Ministry of Education, Youth and Sports of the Czech Republic. We acknowledge Vladimir Sechovsk\'y, Jan Prokle\v{s}ka, Martin \v{Z}onda and Mike Zhitomirsky for insightful discussion.
\end{acknowledgments}

%

\end{document}